# Connected Vehicle Application Development Platform (CVDeP) for Edge-centric Cyber-Physical Systems

Mhafuzul Islam, Mizanur Rahman, Ph.D., Sakib Mahmud Khan,
Mashrur Chowdhury*, Ph.D., P.E., F. ASCE
*Glenn Department of Civil Engineering, Clemson University, Clemson, SC 29634*

&

Lipika Deka, Ph.D.
*School of Computer Science and Informatics, De Montfort University, Leicester, UK*

**Abstract:** *Connected vehicle (CV) application developers need a development platform to build, test and debug CV applications, such as safety, mobility, and environmental applications, in an edge-centric Cyber-Physical Systems. Our study objective is to develop and evaluate a scalable and secure CV application development platform (CVDeP) that enables the CV application developers to build, test and debug CV applications in real-time. CVDeP ensures that the functional requirements of the CV applications meet the latency requirements imposed by corresponding CV applications. We conducted a case study to evaluate the efficacy of CVDeP using two CV applications (one safety and one mobility application) and validated them through a field evaluation at the Clemson University Connected and Autonomous Vehicle Testbed (CU-CAVT). The analysis outcome proves the efficacy of CVDeP, which satisfies the functional requirements (e.g., latency, throughput) of a CV application while maintaining scalability, and security of the platform and applications.*

*To whom correspondence should be addressed. E-mail: mac@clemson.edu.

## 1 INTRODUCTION

The emerging connected vehicle (CV) environment consists of different kinds of entities, such as On-Board Units (OBUs) capable of sending and receiving messages, Roadside Units (RSUs) are also capable of sending and receiving messages from vehicles as well as communicating with personal devices (e.g., cell phone), sensors (e.g., environmental sensors), and Traffic Management Centers (TMCs). With integrated computing and/or control capabilities, these connected physical components communicate with each other to form a Cyber-Physical System (CPS). Considering a large-scale deployment for such connected vehicle CPS, a concept of 'Edge computing' is introduced as the underlying computing approach because of its potential benefits for enabling reduced communicational latency and increased scalability brought about by bringing resources such as storage, and computational resources closer to the edge and consumers (Lopex et al., 2015), (Grewe et al., 2017). The resultant CPS forms an edge-centric CPS. In the edge-centric CPS, the resources for communication, computation, control, and storage are placed at different edge layers (e.g., mobile edge as a vehicle, fixed edge as a roadside infrastructure, system edge as a backend server or TMC) in a CV environment (Rayamajhi et al., 2017)(Zhu, F., and Ukkusuri, 2018). In an edge-centric CPS, a CV application can be divided into sub-applications where different sub-applications run in different edge layers depending on the requirements of the application.

Architecture Reference for Cooperative and Intelligent Transportation (ARC-IT), which has been developed by the U.S. Department of Transportation (USDOT), have listed and provided guidelines for planning and implementation of over a hundred CV applications for safety, mobility, and environmental benefits (ARC-IT, 2018). For example, 'Traffic Data Collection for Traffic Operations' is a CV application, which uses CV data obtained from OBUs to support traffic operations (ARC-IT, 2018). To develop such CV applications for an edge-centric CPS, developers need a dedicated platform where they can develop, test and debug CV applications.

The Operational Data Environment (ODE) system, which is being developed by Intelligent Transportation Systems Joint Program Office (ITS JPO) (USDOT, 2018), is a real-time data collection and distribution software system that collects, processes and distributes data to different components of the CV environment, such as CVs themselves, personal mobile devices, infrastructure components (e.g., traffic signal) and sensors (e.g., camera, environmental sensor). According to architecture of the ODE, CV application developers can stream data using ODE in real-time, and this system does not provide application developers an opportunity for developing, testing and debugging CV applications. Thus, it is critical to develop an application development platform and evaluate the platform in terms of latency and throughput to satisfy the temporal and spatial requirements of CV applications (Du et al., 2017), (USDOT, 2017). This application development platform should also consider the scalability of platform, and security of the platform and CV applications. To the best of our knowledge, there exists no such CV application development platform that provides an application development platform while considering platform scalability, and platform and application security.

The major challenges for developing a CV application development platform for an edge-centric CPS are to (a) enable developers' to collect, process and distribute data, while running multiple CV applications concurrently in real-time in different edge layers; and (b) ensure security of the platform and application while maintaining the scalability of the platform. Hence, the objective of this study is to develop and evaluate a scalable and secure CV application development platform that handles real-time data from CVs in an edge-centric CPS, and can satisfy the requirements imposed by CV applications. This platform, which we call 'Connected Vehicle Application Development Platform (CVDeP),' has been designed to hide the underlying low-level software, hardware, and associated details by providing access via an abstraction layer. An application programming interface (API) layer will provide developers an easy and secure access to the edge devices. In addition, this platform will ensure the scalability of the edge-centric CPS as the penetration level of CVs and number of fixed and system edges varies. Security of the platform is guaranteed by securing access of the developers to the platform, in addition to maintaining application security. An authentication based access control mechanisms are used to secure the access to the platform, and flow-based security policies are used to monitor the data flow for the CV applications to ensure application security. However, developing security policies for detecting cyber-attacks and identifying related countermeasures are not the focus of our study.

A case study has been conducted to evaluate the efficacy of CVDeP using a safety application (i.e., Forward Collision Warning) and a mobility application (i.e., Traffic Data collection for Traffic Operation). These applications were developed and evaluated in CVDeP and later validated in a real-world edge-centric Clemson University Connected and Autonomous Vehicle Testbed (CU-CAVT), which is located at Clemson, South Carolina. 'Forward collision warning' (ARC-IT, 2018) application has been selected as it is a fundamental application for Vehicle-to-Vehicle or V2V safety. Similarly, 'Traffic Data collection for Traffic Operation (ARC-IT, 2018)' application has been selected for the case study, because this application supports many other Vehicle-to-Infrastructure or V2I safety and mobility applications, such as cooperative adaptive cruise control, incident detection and implementation of localized operational strategies (e.g., altering signal timing based on traffic flows, freeway speed harmonization, optimization of ramp metering rates). Using these CV applications, the efficacy of the CVDeP was evaluated using two measures of effectiveness (i.e., latency and throughput).

The remainder of the paper is organized as follows. Related work, which studied the CV application development requirements, real-time data sharing methods, real-time CV data sharing platform, and access control to edge devices and application security are discussed in Section 2. The research method is presented in Section 3. The architecture of an edge-centric CPS for connected vehicles and CVDeP are presented in Section 4. Section 5 presents the implementation of CVDeP for an edge-centric CPS followed by the evaluation of CVDeP in Section 6. Conclusions and future works are discussed in Section 7.

## 2  RELATED WORK

In order to develop the CVDeP that uses real-time CV data, we reviewed existing works related to the CV applications development criteria, and real-time CV data sharing strategies and platform as well as their limitations, developer access control, and application security.

### 2.1 CV application development requirements

Any CV application is bounded by the time and space requirements for providing the desired service (Karagiannis et al., 2011). If CV data are not received within the temporal and spatial threshold as required by specific CV applications, CV data will not have any efficacy for real time CV application. The Michigan connected vehicle testbed 'Proof of Concept Test Report' categorized CV data by time and space contexts (Fehr et al., 2018). While streaming data, timestamp information and location should be included in the CV data as such data are included in the Basic Safety Message (BSM) sets, and they support data validity checks. In addition, data disseminated by the application development platform must be consistent and error-free (Agmon and Ahituv, 1987).

External application developers may require two kinds of data depending on the application, namely real-time disaggregated data and aggregated data. For example, applications such as incident detection applications require real-time disaggregated data for running and testing of algorithms (Du et al., 2017), thus making it necessary for the platform to provide such data. On the other hand, applications such as those that provide 'Queue Warning' after every 5 minutes (Balke et al., 2014) may not require the raw data, but an aggregated data is sufficient. Considering the CV applications that require data from multiple sources (e.g., OBUs, RSUs), a CV environment is considered to be one of the largest distributed networks in the near future (Qian et al., 2008). As the size of the CV transportation network grows (e.g., number of vehicles, sensors, roadside infrastructure), the demand for data will also increase (Baker et al., 2016). Thus, a platform for CV application development needs to be designed in such a way so that it can handle a high demand of data without compromising the quality of service (in terms of temporal and spatial requirements) for the CV applications. Thus, in providing the data to the users, CVDeP needs to meet the application requirement in terms of latency and throughput, and must be capable of handling the scalability issues with increasing number of connected vehicles, sensors and roadside infrastructures.

## 2.2 Real-time data sharing methods

In the context of the Internet of Things (IoT), a number of protocols and methods have been proven to be effective for real-time data sharing. Considering the limitations of processing capabilities, memory or/and storage, and communication bandwidth within the IoT environment, different strategies have been proposed, such as publish-subscribe or broker-based system, and Websocket-based system (Yasumoto et al., 2016). Given that a CV environment is essentially a part of an IoT environment and real-time data sharing is an important goal, the same methods from the IoT domain can be adopted for real-time data sharing in an edge-centric CV environment. RabbitMQ (RabbitMQ, 2017), ActiveMQ (ActiveMQ, 2017), Redis Pub/Sub (Redis, 2017), and Kafka (Apache Kafka, 2017) are the popular platforms that use the publish-subscribe (Broker) based system for real-time data sharing. Websocket (Websocket, 2017) and IPv6 over Low-Power Wireless Personal Area Network (6LowPAN) (Shelby and Bormann, 2010) are other available alternative methods. In our study, we have selected the broker-based system over other systems (e.g., Websocket, 6LowPAN) for real-time data sharing due to its capability of decoupling data sender and receiver, and low computation and bandwidth usage (Chowdhury et al., 2017). Also, among all broker systems, Kafka is selected as a candidate to develop the application development platform for the edge-centric CPS as it shows superior performance in terms of throughput and scalability compared to other popular broker-based systems (e.g., ActiveMQ, RabbitMQ) (Kreps et. al., 2011), (Du et al., 2017).

## 2.3 Real-time CV data sharing platform

Very few research efforts have been conducted to share real-time CV data with multiple application developers who do not have direct access to CV testbed. Currently, the USDOT provides a real-time data sharing platform, named 'ITS Public Data Hub', through which existing CV Testbeds (e.g., New York, Wyoming and Tampa CV Testbed) can submit and share their CV research data. The purpose of the ITS Public Data Hub is to support research, analysis, application development and testing (Open Data Portal, 2018). However, there are a few limitations on the Public Data Hub platform. One of the major limitations of ITS Public Data Hub is that most of the data published in the Public Data Hub are archived data. Real-time data sharing for connected vehicle environment is still at a prototype level (Intelligent Transportation Systems, 2017). On the other hand, ITS JPO (USDOT, 2017) has developed a real-time data acquisition and distribution software system called ODE. The ODE can support collection and distribution of the data to devices', such as CVs, personal mobile devices, infrastructure components (e.g., traffic signals), and sensors. However, the ODE is primarily focused on data acquisition and distribution systems instead of providing an application development platform for CV application developer who can access to the platform and run their applications.

## 2.4 Access control and application security

Security is one of the major concerns in deploying the CV applications because of the vulnerability and safety critical aspect of connected transportation systems (El Zarki et al., 2015), (Karagiannis et al., 2011), (Raw et al., 2013), (Wang et al., 2018). The USDOT proposed a security concept 'Security Credential Management System (SCMS)' to ensure privacy and integrity in a CV system that includes application security. The data shared between the applications and edge devices need to be secured and we need to maintain data confidentiality, integrity, and availability (ARC-IT, 2018). One way to protect the data from unwanted user access is to authenticate user information before sharing the streaming data. In SCMS, the fixed edges (e.g., RSUs) will provide a certificate to the application, which can be used by the application for message exchange (Whyte et al., 2013), (Ahmed-Zaid et al., 2011). Registration Authority (RA) and Certificate Authority (CA) were considered for providing the certificate. RA verifies the user request and checks digital signature. CA issues a new digital certificate or renews a certificate.

In SCMS, the application security is permission-based, which does not check data flow between the authorized components (e.g., data consumer, data producer). In the permission-based control system, the assumption is that the authorized application will not be abused by the given permission. To address the limitation of the permission-based system, we need a dynamic or static taint analysis (Russello et al., 2012). An example of static taint analysis as follows: if an application request a service , check the service request based on policies and perform an annotation with security labels (e.g., benign, or malicious request or data). Unfortunately, taint analysis can encounter difficulty with concurrency (i.e., the data is tainted by two or more applications simultaneously) and implicit flows (e.g., information leak by observing the pattern of output values) (Sarwar et al., 2013). Specialized hardware (e.g., hardware with the type-safe runtime, the processor having Embedded Trace Macrocell) can solve these limitations; however, additional cost and significant computational overhead will be added (Paupore et al., 2015) to implement this system. To reduce such cost and computational overhead, a flow-based control system has been developed (Fernandes et al., 2016). The security mechanism developed by Fernandes et al. (2016) is aimed at IoT frameworks. As the flow-based control system has not studied yet for a CV environment, this flow-based control model can be adopted for CV environment for a CV application security. The flow-based model taints the data using '<*source, sink*>' label to keep track of the data flow between different CV applications. Sensitive data (e.g., driver identity, vehicle id) are removed or modified while the data are delivered to CV applications to ensure the privacy. Another study by (Islam et al., 2018) proposed a V2I application security platform, called CVGuard, based on a set of security policies and rules. As they are also focused on the V2I application security, the concept can be adapted to protect CV applications while the application is developed in different edges (e.g., fixed edges, system edges). In our study, we have implemented a security module to control access, certificate exchange mechanism like SCMS, as well as application security based on data flow policies developed by Islam et al., (2018).

## 3 CONNECTED VEHICLE APPLICATION DEVELOPMENT PLATFORM (CVDeP)

Our research approach, which is illustrated in Figure 1, in this study includes conceptual development, implementation and evaluation of CVDeP. Edge computation, communication and security are the key components for developing a connected vehicle application development platform in an edge-centric CPS. Based on these required components, a CVDeP architecture is developed including application management platform and application development graphical interface. Computation, communication, security and graphical interface modules are then implemented based on the architecture of CVDeP. After that, we evaluate the CVDeP using safety and mobility applications at two different phases: i) simulation evaluation and; ii) field validation. We explain the experimental set-up, experiment scenarios and CV applications, for the evaluation, in detail at Section 6. In the following sections, we have presented the above-mentioned research approach in detail for developing our CVDeP.

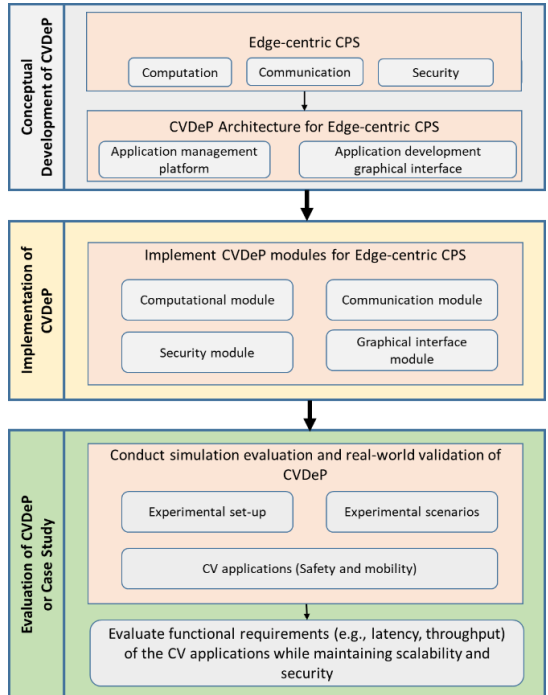

**Figure 1** CVDeP Development and Evaluation Framework

## 4 CONCEPTUAL DEVELOPMENT OF CVDeP

In this section, we describe the conceptual development of CVDeP for an edge-centric CPS. First, we identify the necessary components of an edge-centric system, and then define the architecture of CVDeP based on the edge-centric CPS.

### 4.1 Edge-centric Cyber-Physical System

For an edge-centric CPS, we have developed CVDeP to support CV application developers. The primary concept behind edge-centric computing is decentralization by distributing the resources, such as computing and storage, from the central cloud nodes to the edges and closer to the consumers. This physical proximity is envisioned to reduce latency and the distributed architecture aims at increasing scalability. The edge-centric CPS as shown in Figure 2 for CV systems consist of three edge layers: i) mobile edge (e.g., On-board sensors); ii) fixed edge (e.g., roadside transportation infrastructure); and iii) system edge (e.g., backend server at Traffic Management Center (TMC)) (Rayamajhi et al., 2016). This hierarchical cyber-physical system architecture, which is presented in Figure 2, can address complexity and scale issues of CV systems. A system edge is a single end-point for a cluster of fixed edges. A fixed edge includes a general-purpose processor (i.e., application development device) and a Dedicated Short Range Communication (DSRC) based RSU. A fixed edge can communicate with the mobile edges using DSRC and communicate with system edge using Optical Fiber/Wi-Fi. Fixed edge can be extended to support a video camera and other sensing devices, such as weather sensors and GPS sensors. CVs participating in our system will be acting as mobile edges, and be equipped with a DSRC-based OBU. Fixed edges are connected to a system edge that can effectively serve as a backend resource. Mobile edges (Edge layer 1) can exchange data with fixed edges (Edge layer 2) and system edges (Edge layer 3) using DSRC and LTE/Wi-Fi communication, respectively as shown in Figure 2.

To ensure reliable CV data delivery via different communication mediums, heterogeneous wireless network (HetNet) service is utilized in each layer of the edge-centric CPS. HetNet is an important enabler of edge-centric CPS for CVs as it decides the available best communication medium to be used for a particular application depending on the feasibility, accessibility, and data delivery requirements (i.e., temporal and spatial requirements) of a CV application (Du et al., 2017), (Dey et al., 2016). Real world deployments would cover large geographical areas where they would require more than one network technology, which includes DSRC, Wi-Fi, LTE and optical fiber, to support communication needs. In edge-centric CPS for CV, choice of communication networks depends not only on its placement of components (location of mobile edges from the fixed edge and system edge), but also on requirements necessitated by CV applications. In the edge-centric CPS architecture adopted for the CVDeP, the fixed edge can support more than one networking technology, and are designed to be equipped with DSRC Radios, Wi-Fi hotspots, Cellular 3GPP, LTE networks or optical fiber networks. They are also designed to support a wide range of applications, and sensors, such as weather monitoring sensors and video cameras. A publish-subscribe messaging system can be employed at each edge to support data transfer among the connected edges (e.g. mobile edge, fixed edge, system edge) (Chowdhury et al., 2018). A message can be transferred by a tag or a topic, and message content of that topic are routed through different communication networks, and reach a specific edge layer based on the subscription of topics.

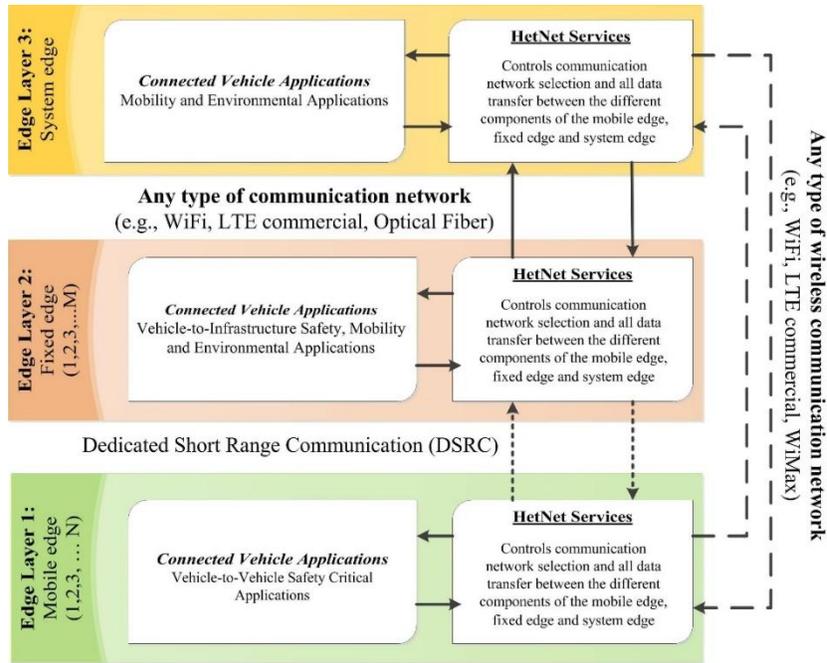

**Figure 2** Edge-centric CPS architecture for a connected vehicle environment

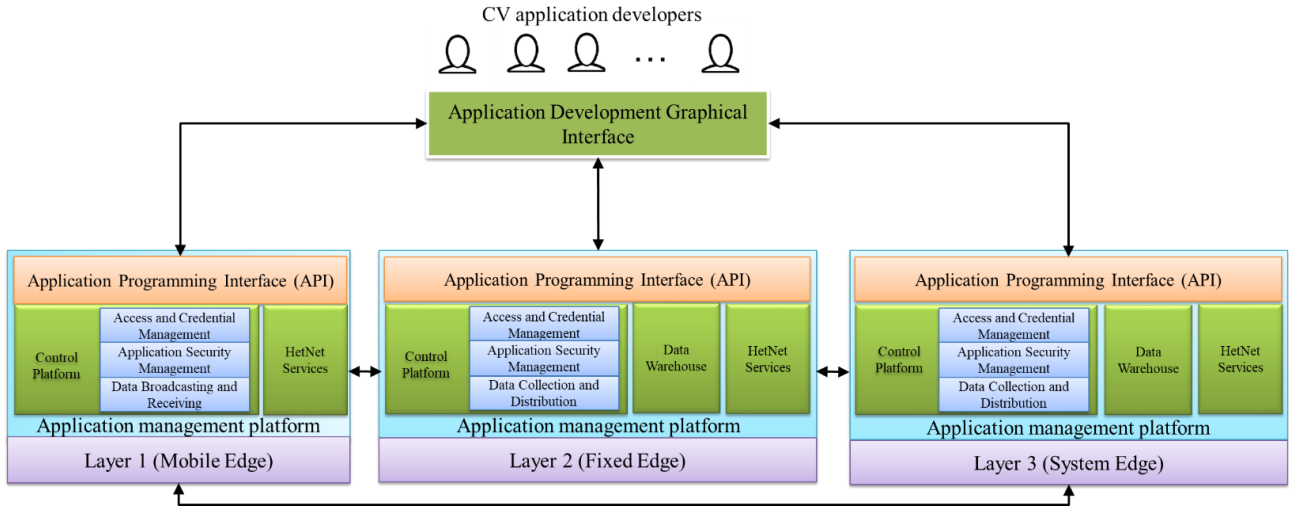

**Figure 3** CVDeP architecture for an edge-centric CPS

### 4.2 CVDeP Architecture

In an edge-centric CPS for CVs, each component generates different types of data. For example, OBUs installed in a vehicle (i.e., mobile edge) broadcast BSMs, which contain the vehicles' information, such as location, speed, direction, acceleration, and braking status (Kenney, 2011). The fixed edge (i.e., RSU with an additional edge device that has computational power) collects data from the OBUs and acts a primary gateway to transfer data from CVs to the transportation infrastructures (e.g., system edge, which could represent a TMC). For developing a CV application, developers need to interact with all of the layers mentioned above. Hence, edge layers are accessed through an Application Programming Interface (API), which provides a way for the CV application developers to interact with the different edge layers. Figure 3 illustrates the architecture of our CVDeP for an edge-centric CPS which comprised of the two following components: 1) application management platform, and 2) application development graphical interface.

#### 4.2.1 Application management platform

Application management platform is responsible for the selection of the appropriate communication medium of an application, and data collection, storage and distribution, while ensuring the security of the platform by providing secured access control and security management of the CV applications. As presented in Figure 3, application management platform resides in between the application development graphical interface and the underlying CV components (i.e., each edge) of the edge-centric CPS. Application developers interact with management platform through an API. The application management platform is made up of the following components: (i) control platform; (ii) data warehouse; and (iii) HetNet service.

*4.2.1.1 Control platform:* The control platform supports four types of operations: (1) data broadcasting to and receiving from the various mobile edge devices; (2) data collection and distribution (for fixed edges and system edges); (3) access control and credential management; and (4) application security management. Edge devices on an edge-centric CPS continuously exchange data between different edges. The data broadcasting and receiving module in the mobile edges handles the continuous data exchange between mobile edges and other edges (i.e., the system edge and the fixed edge). This module continuously provides BSMs to the application developers that can be used to develop CV applications. On the other hand, the data collection and distribution module in fixed edges and system edges is responsible to gather and distribute data to and from mobile edges, fixed edges, and system edge in real-time. Both the broadcasting-receiving module and collection-distribution module can be used by the developer to develop any type of CV applications. After the access control and credential management component are activated, authenticated application developers can access, gather and visualize real-time streaming data generated from different components of each edge layer. In addition, application security management module is responsible for monitoring the data flow and securing the application using security policies.

*4.2.1.2 Data warehouse:* The data warehouse stores the data generated from different edge devices, sensors, and applications deployed in the edge layers. It is a distributed storage system which resides in the fixed edge and the system edge. The purpose of the data warehouse is to store and provide necessary information that is needed by the developers and/or edge layers for any application's needs. As a mobile edge is limited by computation power and storage size, we do not include a data warehouse in mobile edges. In fixed edges and system edges, the structure of the data warehouse is such that it can support and store both structured (e.g., GPS data) and unstructured data (e.g., text and images).

*4.2.1.3 HetNet services:* HetNet services decide the available best communication medium to be used for a particular application. Developers will provide temporal and spatial requirements of an application to HetNet services through application development graphical interface, and then HetNet services create an abstraction layer for the developers on top of the internal communication networks. For example, HetNet selects DSRC, which is low latency communication medium, from the available communication options to satisfy the requirement of a V2V and V2I safety applications. While the application is running in edge devices, CVDeP will provide the HetNet metadata (e.g., available communication mediums such as DSRC, LTE, and Wi-Fi, and their average, maximum, and minimum transmission latency) for evaluating the performance of the application.

### 4.2.2 Application development graphical interface

Application developers can access the underlying edge devices of the edge-centric CPS using a graphical user interface (GUI). A block diagram of the GUI is presented in Figure 4. One can develop and deploy any CV application directly on the edge-centric CPS via the GUI of the application development graphical interface. Based on the requirements of a CV application, interface access rights and available services (e.g., HetNet services, Data storage service) of the platform, application developers can access to the different types of data (e.g., real-time and historical) through API in each layer. Using this API in each layer, application developers can also request any specific data for a specific application purpose. For example, developers can request the data from data warehouse to predict the future roadway traffic condition. Application development graphical interface will provide an interactive platform to the developers to build their own applications and test these applications by requesting both real-time data from CVs and other sensors, and historical data from the data warehouse from fixed and system edges.

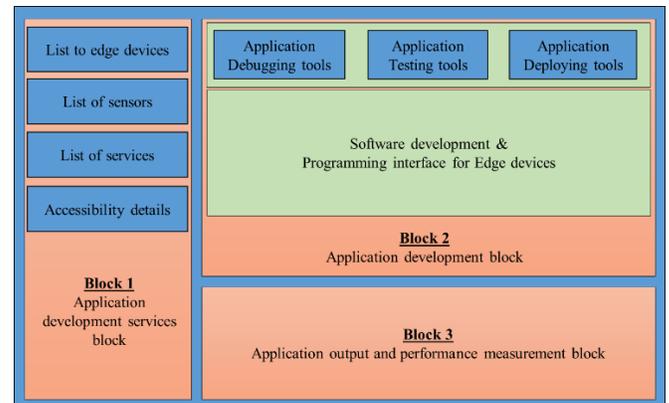

**Figure 4** Block diagram of the CVDeP application development graphical interface

As shown in Figure 4, the development graphical interface is divided into three blocks. Using the functionalities of block 1, a developer can connect to the edge devices through an authentication procedure using the accessibility details (i.e., username and password) provided by the platform. After the authentication procedure, developers will also be aware of the list of available edge devices (e.g., number and type of edge devices), services (e.g., HetNet service, applications' output), and sensors (e.g., GPS, Camera, environmental sensors) in each device. Then, using block 2, they can develop, deploy, test and debug the application in edge devices using the provided tools, and software development and programming interface. After deploying an application, developers can observe the output and performance of an application and save these data through the application output and performance measurement block (block 3).

## 5 IMPLEMENTATION OF CVDeP

The implementation of CVDeP components (namely the application management platform and application development graphical interface) and their sub-components (namely the control platform, data warehouse and HetNet services) are discussed in the following sub-sections.

### 5.1 Application management platform

The implementation details of the control platform, HetNet services, and data warehouse are presented in the following sub-sections.

*5. 1.1 Control Platform.* The control platform contains four modules depending on whether the edge device is a mobile, fixed or system edge. Implementation overviews of these modules are as follows:

- *Data broadcasting and receiving.* The data broadcasting and receiving module is developed for only the mobile edge devices, where it is responsible for generating BSMs and receiving the BSMs from other mobile edges. In our implementation, each mobile edge is broadcasting BSMs at a default rate of 10Hz and each BSM contains necessary attributes for safety applications (e.g., position, speed, and direction) of corresponding mobile edge (Park, 2014, USDOT, 2013), (Kenney, 2011). In addition, each mobile edge is receiving the BSMs from all other mobile edges within their communication range.

- *Data collection and distribution.* Data collection and distribution system is the core part for fixed and system edges of CVDeP. We have selected Kafka as a broker-based system data collection and distribution systems because of the following efficacies (Kreps et al., 2011): 1) high throughput; 2) low latency; 3) reliability of data delivery, and 4) scalability. In a publish-subscribe based broker-system, data producers (e.g., mobile edges, applications) produce and publish data to the broker, whereas the data consumers (e.g., fixed edge, applications) subscribe and consume the data available at the broker. By tagging individual data elements with labels/topics, producers can produce data for a particular topic and consumers can subscribe data of that topic. Brokers receive data from producers and immediately make the data available for consumers to consume. As a result, producers and consumers can generate and consume data in an asynchronous and independent manner.

*Access control and credential management system. The* access control and credential management module ensures that only authorized users have access to the CVDeP services. Developers are authenticated via a login interface before given access to the edge-centric CPS testbed components. A permission-based access control is implemented by providing access rights to application specific data and services (e.g., access to the BSMs, access sensors data, access to the data warehouse) like Android application system where permission are written in a manifest file prior to a developers develops the Android application (Felt et al., 2012). On the other hand, the Credential Management System (CMS) was implemented based on the Public Key Infrastructure (PKI), which takes care of public key exchange that is needed for encrypting and authenticating data using a digital signature. CMS is built in such a way that can replicate the functionalities of SCMS proposed by USDOT (Whyte et al., 2013). As shown in Figure 6, consumers and producers are provided a certificate trough an API by CMS before any data exchange. Then they use the certificate to encrypt their message and send the message to broker-based data collection and distribution systems. We have followed the assumptions by the National Highway Traffic Safety Administration (NHTSA) connected vehicle pilot program where V2V communications is secure, but not encrypted, and V2I communication is both secure and encrypted (Weil, 2017).

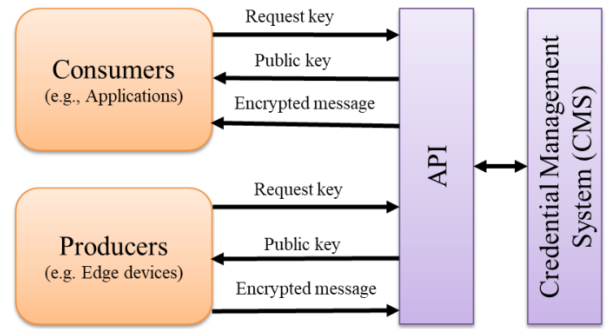

**Figure 6** Implementation of access and credential management architecture for CVDeP

- *Application security management.* The flow-based control module as proposed by Fernandes et al (2016), is included in the data collection and distribution systems to ensure application security. Initially, all the consumers and producers need to be authenticated (Action 1 (A1) and Action 2 (A2) respectively in Figure 7) to produce and consume the message. Then they are allowed to produce (A3 in Figure 7) and consume (A6 in Figure 7) data from the Data collection and distribution module. In the security module, trusted API and quarantine module checks the flow policies (A4 and A5) and deliver the data (A6) to the appropriate consumers (e.g., the consumers who are authenticated and subscribed to a particular topic). As shown in Figure 7, producers and consumers communicate with the data collection and distribution module via a trusted API. This trusted API removes any sensitive information (e.g., drivers identify, vehicle ID etc.). Moreover, this trusted API enforces the flow policies among the applications. Using these flow policies, the application security can be ensured. In our study, we have implemented the flow policies using '<*source, sink*>' tracking as described in (Fernandes et al., 2016) in which source is the producer of the data and sink is the intended recipient of that data.

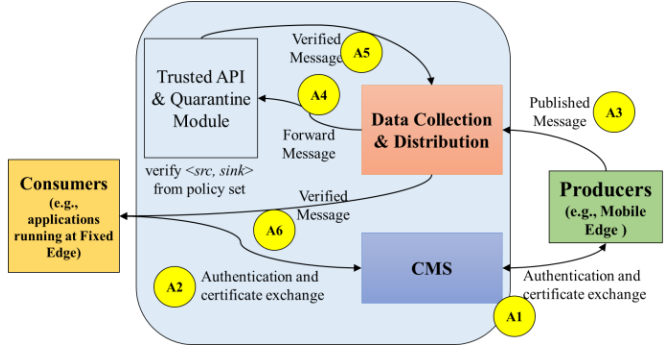

**Figure 7** Implementation of application security module with data collection and distribution systems

*5.1.2 Data warehouse*: In a CV environment, a massive amount of data will be generated from different type of devices and applications on different edge layers. Data from different sources can have different data structures, such as structured, semi-structured and unstructured. A structured data has a strict tabular format whose column size and attributes of each entity are defined. Examples of structured data include any data that can be stored in delimited formats, spreadsheets, or SQL tables, whose columns are defined. A semi-structured data includes data whose fields are defined but organized in a hierarchical manner. Examples include data stored in Extensible Markup Language (XML) or JavaScript Object Notation (JSON) formats. Unstructured data, such as pictures, videos, and textual data, do not have any structural organization associated with the data itself. In our implementation, to support structured, semi-structured, as well as unstructured data, we have used MySQL for structured data and NoSQL for semi-structured and unstructured data. With the structured, semi-structured and unstructured data together produces a huge amount of data in terms of volume. Realistically, we do not need to store all the raw data in their original format. As a result, a lambda infrastructure (e.g., Amazon web service), which is designed to handle data in massive quantities using batch processing, can help to reduce and compress historical data for subsequent batch processes.

*5.1.3 HetNet services*: As mentioned in sub-section 4.2.1.3, HetNet manages the underlying communication networks in the edge-centric CPS environment. The HetNet services are implemented in the network layer to manage the connectivity using the available communication mediums between the available edge devices. HetNet services are embedded in the edge-centric CPS and the best available communication medium is decided based on measured network statistics (e.g., end-to-end metrics, which are maximum, minimum and average transmission latencies, packet loss, and signal strength). Figure 8 illustrates the implementation of HetNet in

CVDeP. HetNet services monitor the existing networks and select a network suitable for an application and the control platform controls the flow of the application data. The decision of selecting among available communication medium is done based on the application requirements set by the developers and available communication network statistics.

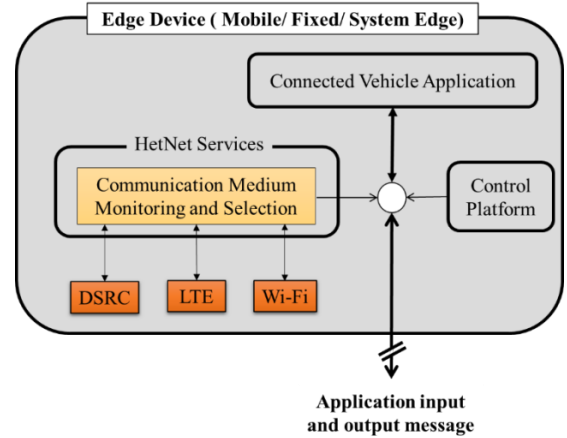

**Figure 8** Implementation of HetNet services

In our HetNet services implementation, the discovery or searching of communication mediums and their network statistics are measured in the background asynchronously. An application is agnostic of the communication mediums and the decision of the medium to use for transmitting and receiving data is decided by the HetNet services while control platform functionalities are involved throughout the process as described in the previous sections. We have added the metadata support layer in the HetNet to provide metadata to the developers that can support them to develop their applications. Through this metadata layer, developers will be able to observe the communication attributes, such as signal strength, bandwidth utilization and data loss. A script running in CVDeP provides this information to the developers, and developers can evaluate the effect of communication medium on the performance of an application.

## 5.2 Application development graphical interface

The application development graphical interface is developed as a desktop application in C# (C sharp) as illustrated in Figure 5. The implemented features of this application development graphical interface are described in further detail in section 4.2.2. Currently, the software has only been developed for the windows operating systems as a proof of concept.

## 6 EVALUATION OF CVDeP: CASE STUDY

This section provides a description of the case study area and case study with a safety and a mobility application, which are developed in CVDeP to prove the efficacy of CVDeP. For our case study, we have developed 'Forward Collision Warning (FCW)' as a safety application and 'Traffic Data Collection for Traffic Operations' as a mobility application (ARC-IT, 2018) using CVDeP. Then, to prove the efficacy of CVDeP, these applications are evaluated in our emulated environment in CVDeP and real-world Clemson University-Connected and Autonomous Vehicle Testbed (CU-CAVT) at Clemson, SC (Chowdhury et al., 2018).

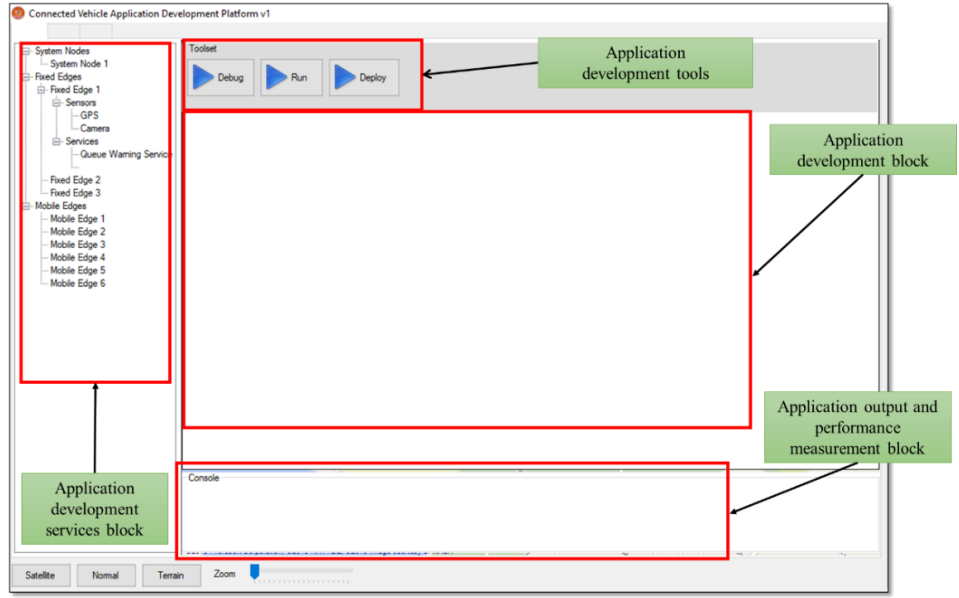

**Figure 5** Implementation of application development graphical interface

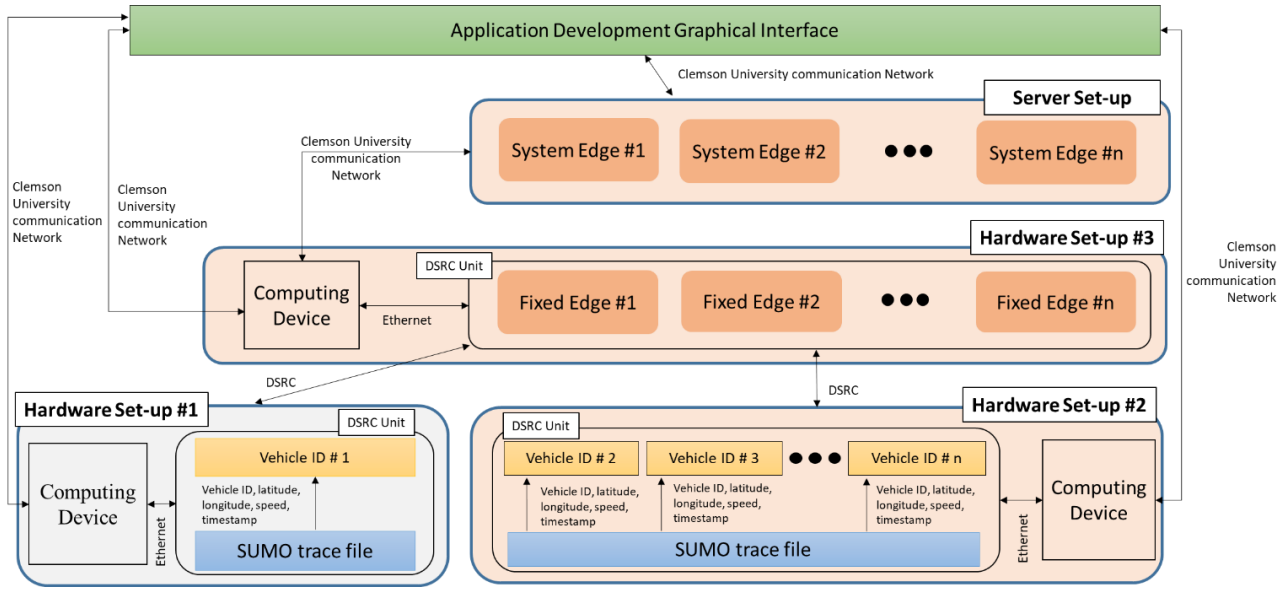

**Figure 9** CVDeP environment for emulating edge-centric CPS

## 6.1 Experimental Setup for CVDeP and CU-CAVT

In this section, we will describe emulated connected vehicle environment setup in CVDeP and CU-CAVT in detail.

### 6.1.1 CVDeP

Developer can develop and evaluate the performance of the developed CV applications in the CVDeP environment. In this environment, developer will have dedicated hardware to emulate the real world edge-centric CPS for CVs. As shown in Figure 9, developer can emulate mobile edges using Hardware Set-up #1 and #2, and fixed edges using Hardware Set-up #3, Whereas system edges has been set-up in a dedicated server in Clemson University. Each Hardware set-up (#1, #2, and #3) consists of one DSRC unit to send and receive the DSRC messages, and computing device for computation as well as communication purpose (see Figure 9). Hardware set-up #1 is used for developing the safety application where the safety application is intended to develop and test (see section 6.2.1 for more explanation). Hardware Set-up #2 is used for emulating other mobile edges for safety application. For mobility and environmental application, only Hardware Set-up

#2 (see Figure 9) can be used for emulating mobile edges (see section 6.3.1 for more explanation). Hardware setup #3 (see Figure 9) is used for creating any number of fixed edges where the location of fixed edges are defined by the developers through CVDeP interface. A dedicated server computer, which is available for server set-up, located in Clemson University is intended for creating system edge instances. In this emulated edge-centric CPS, mobile edges and fixed edges communicate with each other using DSRC, and fixed edge and system edge communicate using the Clemson University communication network, which includes optical fiber and Wi-Fi. In addition, developers can configure the number of edges in each layer as required by the application. To generate the movement data of mobile edges, the simulated movement of the mobile edges are exported from the 'Simulation of Urban Mobility (SUMO) (SUMO, 2018)', which is a microscopic traffic simulator software, using a SUMO trace file. Using this SUMO trace file, developers can create any roadway environment, and generate any number of emulated vehicles and their corresponding BSMs. A program running in mobile edges read that trace file and generate BSMs for each vehicle. Then, these BSMs are broadcasted using DSRC for each vehicle. The fixed edges will receive BSMs only within its communication range of a fixed edge, which is defined by the developers. According to Figure 9, a developer can access the edges through CVDeP Interface in order to develop and evaluate the performance of the developed application.

### 6.1.2 CU-CAVT

CU-CAVT is the layered edge-centric CPS based testbed for CVs deployed in the Perimeter Road at Clemson, South Carolina (as shown in Figure 10) (Chowdhury et al., 2018). CU-CAVT has three fixed edges, which are deployed along the Perimeter Road in Clemson, South Carolina, and one system edge is deployed as the backend server. The backend server is located at Clemson University and connected to the Clemson University network. Two of the fixed edges are connected to the Clemson University Network via optical fiber link and one fixed edge is connected to Clemson University network using Wi-Fi link. Each fixed edge has its own DSRC radio to communicate with mobile edges. Each mobile edge (primarily OBUs on vehicles) is equipped with wireless communication devices such as DSRC, LTE and Wi-Fi. In addition, HetNet services are available in the CU-CAVT for Mobile and fixed edges.

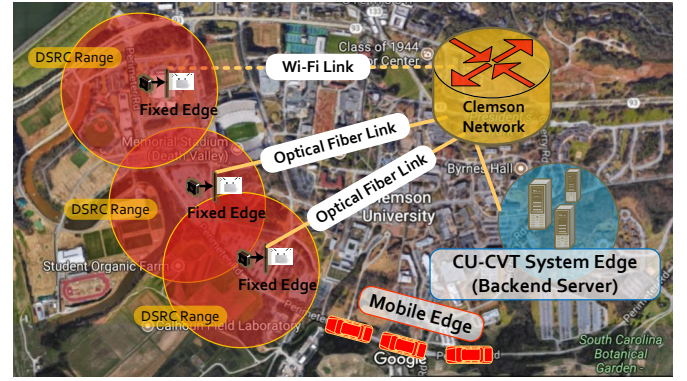

**Figure 10** Clemson University Connected and Autonomous Vehicle Testbed (CU-CAVT).

### 6.2 Case Study 1: Safety Application

For developing a safety application using CVDeP, we have selected FCW application as our candidate application. A study by Xiang et al. (2014) proposed a vehicle kinematics (VK) model for rear-end collision warning application using DSRC communication. Based on their VK model, FCW application generates collision warning when two vehicles are closer than the defined safe distance. Equation (1) shows a modified version of this FCW application as below:

$$D_w = \frac{(V_o - V_t)^2}{2 * a_{moderate}} + d \qquad (1)$$

Where $D_w$ is the distance threshold for collision warning, $V_o$ is the preceding vehicle's speed, $V_t$ is the follower/target vehicle's speed. The follower/target vehicle is the vehicle where the FCW application is intended to run in reality; $d$ is the average length of the preceding and following vehicles, and $a_{moderate}$ is set to 11 ft/s$^2$ following the SUMO configuration. In the following subsections, we present the experimental set-up, evaluation scenarios and evaluation results for the FCW application.

### 6.2.1 Experimental setup

To develop and evaluate FCW application in CVDeP, we have set-up two separate hardware (Hardware set-up #1 and Hardware set-up #2) acting as mobile edges as shown in Figure 11. Hardware set-up #1 acts as the target mobile edge, where the safety application will run and Hardware set-up #2 is used to generate data of other mobile edges. These mobile edges can communicate with each other using DSRC. As described in section 6.1.1, developers can generate BSMs to evaluate their safety application and edge devices (mobile and fixed edges) are located in Clemson University Network in our experimental setup. For the target mobile edge for application development, Control platform and HetNet services are implemented and included. While we evaluate the FCW application, the application security is ensured by control platform, and HetNet services in mobile edge decide which communication medium will be used to send the FCW messages to a fixed edge. In our evaluation of safety application, we are considering only V2V safety application (FCW application), but with our experimental setup developers can also develop V2I safety application using the DSRC communication between fixed edges and mobile edges.

### *6.2.2 Evaluation Scenarios*

We create two evaluation scenarios for evaluating the CVDeP as a safety application development platform (as shown in Table 1). In the scenario 1, the preceding vehicles (Hardware Set-up #2 in Figure 11), and follower or target vehicle (Hardware Set-up #1 in Figure 11) is moving in the same lane with 20 mph and 30 mph, respectively. In scenario 2, both front and follower vehicle are moving with 30 mph and the front vehicle stops suddenly. In both scenarios, FCW application is deployed in the follower vehicle, and forward collision warning is generated based on the comparison between calculated safety distance (using Equation 1) and the distance between two vehicles using real-time GPS data. To evaluate the performance of the application we have considered data delivery latency as a measure of effectiveness. In this context, latency is the time when data was generated by a mobile edge to the time when application produced FCW message in the follower vehicle. Here, latency includes both network latency and computational latency.

#### Table 1
Evaluation scenarios for FCW application

| Evaluation Scenario | Description of evaluation scenario |
|---|---|
| **1** | Preceding vehicle moving at a maximum 20 mph and follower or target application development vehicle approaching with higher speed with 30 mph speed |
| **2** | Both the preceding and follower or target application development vehicles are moving with 30 mph and the front vehicle stopped suddenly |

### *6.2.3 Evaluation in CVDeP*

We have evaluated the safety application, FCW, using the experimental setup as described in the previous section (6.2.1). The application is developed using CVDeP application development graphical interface, and then the application is tested using the two evaluation scenarios presented on Table 1. In the first evaluation scenario, the preceding vehicle is moving at a slower speed than the follower vehicle's speed. In the second evaluation scenario, application in the follower vehicle was activated when the distance between vehicles was below the safe distance. Table 2 provides the summary of latency recorded both from evaluation scenarios using CVDeP. For the evaluation of FCW application in CVDeP, we have taken the data sample of 200s containing 4000 BSMs from 2 mobile edges to calculate the maximum, minimum, and average latency. The average latency is 16 ms for both evaluation scenarios 1 and 2, respectively. However, the recorded maximum latencies were 95 ms and 77 ms, which is below the safety application latency requirement (i.e., 200ms (Dey, 2016), (Whyte et al., 2013)). In Table 2, we present the network latency only. The computational latency for running the application is 1.5 ms, which is same for both evaluation scenarios. In addition, this FCW message is sent to fixed edge using the best available communication medium decided by HetNet which takes 0.5ms to decide, given that all communication mediums (LTE, Wi-Fi, and DSRC) are running simultaneously, and HetNet is monitoring these mediums asynchronously.

#### Table 2
Summary of latency for FCW application evaluation using CVDeP

| Evaluation Parameter | Application Security and HetNet Service | Latency | | Latency requirements for Safety Application (Dey et al., 2016), (Ahmed-Zaid et. al., 2011). |
|---|---|---|---|---|
| | | Scenario #1 in CVDeP | Scenario #2 in CVDeP | |
| Maximum Latency | Included | 95 ms | 77 ms | ≤ 200 ms |
| Average Latency | | 16 ms | 16 ms | |
| Minimum Latency | | 2 ms | 2 ms | |

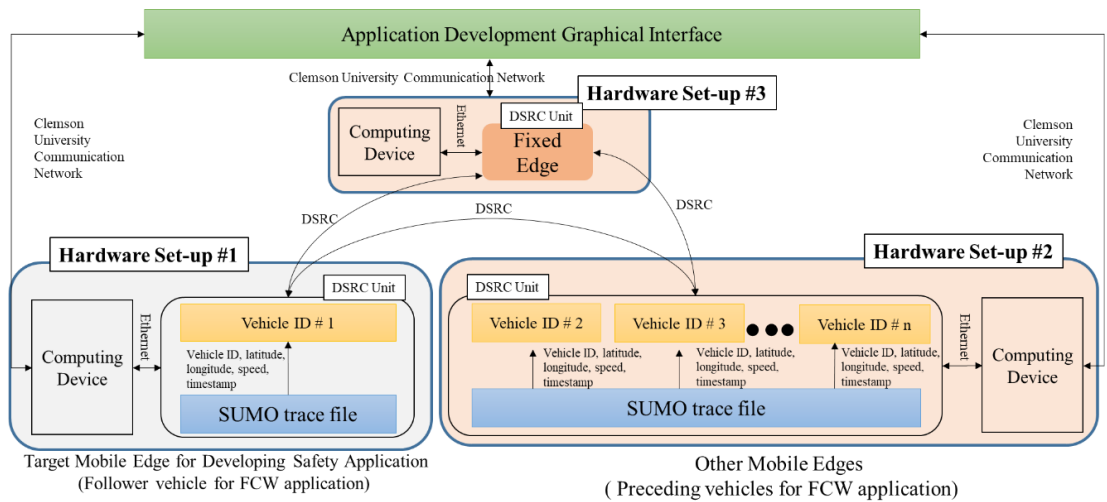

**Figure 11** Experimental setup for safety application

### 6.2.4 Field Validation

For our field evaluation of FCW in CU-CAVT, we followed the similar speed profile in the field experiment for both evaluation scenarios provided in Table 1, and measured the communication latency for the FCW application (as shown in Table 3). Table 3 provides the summary of latency recorded both evaluation scenarios in the field experiment. For field evaluation, same like the evaluation in CVDeP, we have taken the data sample of 200s containing 4000 BSMs from 2 mobile edges to calculate the maximum, minimum, and average latency. The average latency measured is 63 and 49 ms for scenarios 1 and 2 respectively. But the maximum latency recorded for the test is 113 ms and 105 ms, which is still below the safety application latency requirements (i.e., 200ms (Dey et al., 2016; Whyte et al, 2013). In our field experiment, we have observed lower latency than the latency measured in emulated experimental setup because of no environmental effect or propagation loss. As provided in Table 3, we only present the network latency, and we do not present the computational latency for an application which was 2 ms. In both cases (scenario 1 and 2), we can validate that the application developed in emulated CVDeP experimental setup was able to fulfill the application latency requirement (200ms) in the field experiment. Same as before, HetNet services take about 0.5ms time on average to decide the communication medium to use to send the FCW message to upper Edge layers.

### 6.3 Case Study 2: Mobility Application

We have evaluated our CVDeP using 'Traffic Data Collection for Traffic Operations' application. This application collects CVs' data (e.g., BSMs) to support traffic operations, such as incident detection and localized traffic operational strategies

(ARC-IT, 2018). According to this application, it requires to divide the application into two sub-applications: i) Sub-App 1: collect real-time traffic data from mobile edges; and ii) Sub-App 2: collect real-time traffic data from fixed edges (as shown in Figure 12). Sub-application 1 runs in each fixed edge (RSU) and Sub-application 2 runs in the system edge. In the following subsections, we present experimental set-up, evaluation scenarios and evaluation results for the 'Traffic Data Collection for Traffic Operations' application.

**Table 3**
Summary of latency for FCW application evaluation in CU-CAVT

| Evaluation Parameter | Application Security and HetNet Service | Latency | | Latency requirements for Safety Application (Dey et al., 2016), (Ahmed-Zaid et. al., 2011). |
| --- | --- | --- | --- | --- |
| | | Scenario #1 in Field | Scenario # 2 in Field | |
| Maximum Latency | Included | 113 ms | 105 ms | ≤ 200 ms |
| Average Latency | | 63 ms | 49 ms | |
| Minimum Latency | | 2 ms | 3 ms | |

### 6.3.1 Experimental setup

To evaluate our mobility application, we have setup the mobile edges, fixed edges and system edge in the CVDeP environment, and implemented the sub-application of 'Traffic Data Collection for Traffic Operations' as shown in Figure 11.

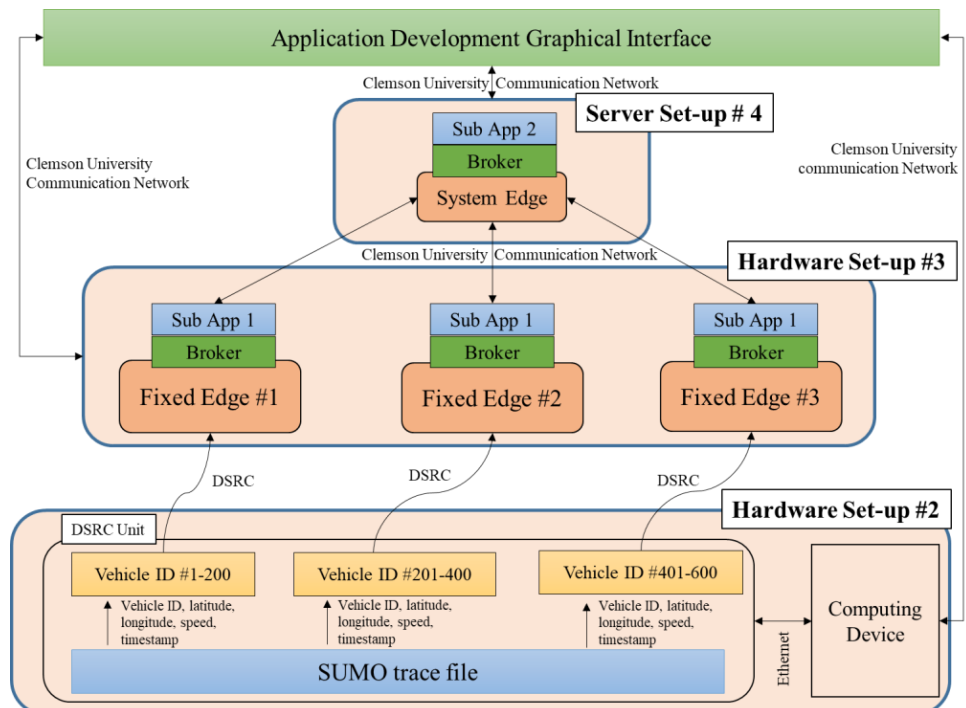

**Figure 12** Experimental setup for mobility application

In CVDeP, hardware set-up #2 is used to emulate mobile edges and to generate BSMs from each mobile edge. Hardware set-up #3 is used to emulate three fixed edges. We have selected the number of fixed edges following the same configuration as CU-CAVT (see Figure 12). Similarly, one system edge has been created in Clemson server. As shown in Figure 12, each sub-application (Sub App 1 and Sub App 2) receive data from Mobile and fixed edges via broker-based data collection and distribution module. Using the same procedure described in section 6.2.1, the road network and movement of the mobile edges on the Perimeter Road in Clemson are simulated using the SUMO tool and a program running in each mobile edge that creates the BSMs at a 10 Hz rate (10 messages per second). Developers interact with the edge devices using CVDeP application development graphical interface.

### 6.3.2 Evaluation Scenarios

We create four different scenarios for evaluating our application development platform varying the number of fixed edges (RSU) and the number of mobile edges as shown in Table 4. In the scenario 1, we have one system edge and one fixed edge with varying number of mobile edges. In the scenario 2, one system edge, multiple fixed edges (RSUs) and 200 mobile edges (200 CVs) for each fixed edge are considered. For evaluation scenario 2, based on fixed edge's coverage, the number of CVs on Perimeter road approaching to the intersection stop line is 200 (maximum number of CVs for four-lane (two lanes in each direction) road during a congested condition according to our traffic volume count). For each scenario, we have evaluated the scalability of the application development platform in terms of data delivery latency and throughput.

### 6.3.3 Evaluation in CVDeP

We evaluate the scalability of our designed CVDeP to ensure the CV application requirements are meet in terms of latency and throughput. The latency is the time difference between the time of data generation at the edge-centric CU-CAVT and the time when the data is received by the user. Data delivery latency requirement for any mobility and environmental applications must be satisfied in order to provide mobility and environmental services. As CVDeP targets to support different mobility and environmental applications, we have considered 1 second (1000 milliseconds) as the latency threshold to deliver the CV data to the developer (Fehr et al., 2014). Also we need to ensure a high throughput (i.e., the data transfer rate) means the high use of the allocated bandwidth. Our platform already fulfilled the spatial requirement of the application, as mobile edges will be within the communication range of fixed edges.

For our evaluation, we implement a data collection and distribution systems (the broker-based system) that is required for real-time application development platform. We evaluate the scalability of the CVDeP considering access and credential management and application security modules with different data collection and distribution systems. Then we compare with latency requirement for the selected CV application. We have used different numbers of mobile edges (i.e., CVs with OBUs), varying from 5 to 200 (See scenario 1 in Table 4), to measure the latency and throughput. As shown in Figure 13 and 14, with the increasing number of mobile edge and fixed edge, the throughput of the broker-based system is linearly increasing and reaches a maximum at 5.2 Mbits/s and 8.4 Mbits/sec, respectively. Higher throughput ensures reliable and scalable services. The broker-based system (e.g., Kafka) uses an asynchronous mode that can collect and distribute data in memory and send them in batches in a single shot (Apache Kafka, 2017). Because of this asynchronous mode and sending data in batch, the broker-based system can ensure high throughput. In the broker-based system, the system adapts the application development platform's throughput as the number of mobile edges and fixed edge increases and thus can handle more data. We observe that CVDeP data collection and distribution system can maintain a lower latency with increasing number of mobile edges (See Figure 15) and fixed edges (See Figure 16). The increment of latency with the broker-based method is negligible for both use case scenarios (scenarios 1 and 2). The reason is that broker-based system uses an intelligent 'sendfile' method with zero-copy optimization (i.e., sending the data directly to the consumer without any buffering or copying to memory) (Apache Kafka, 2017).

**Table 4**

Evaluation scenarios with varying the number of mobile edges and fixed edges

| Scenarios | Data Collection and Distribution System | Access and Credential Management System | Application Security | HetNet Services | Number of System Edge | Number of Fixed Edge | Total Number of Mobile Edge |
|---|---|---|---|---|---|---|---|
| 1 | Broker-based System | Included | Included | Included | 1 | 1 | 5,10,20,30,50,100, 150,200 |
| 2 | Broker-based System | Included | Included | Included | 1 | 1 | 200 |
| | | | | | | 2 | 400 |
| | | | | | | 3 | 600 |

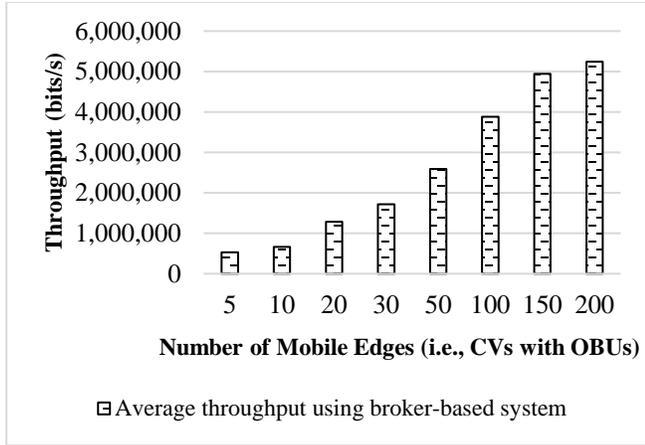

**Figure 13** Comparison of average throughput with different number of mobile edges (i.e., CVs with OBUs) (Scenario 1)

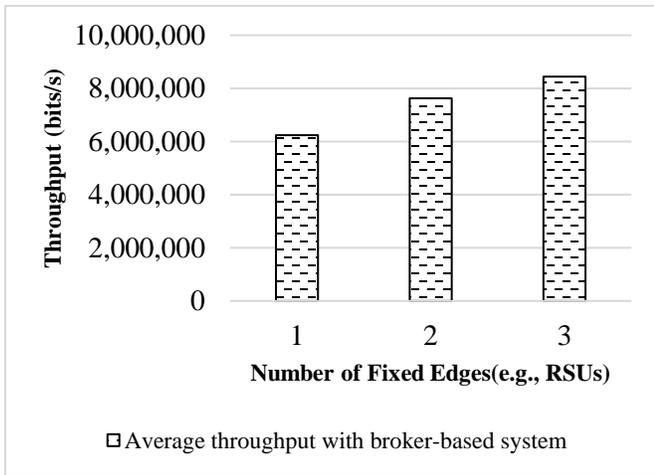

**Figure 14** Comparison of average throughput with different number of fixed edges (i.e., RSUs) (Scenario 2)

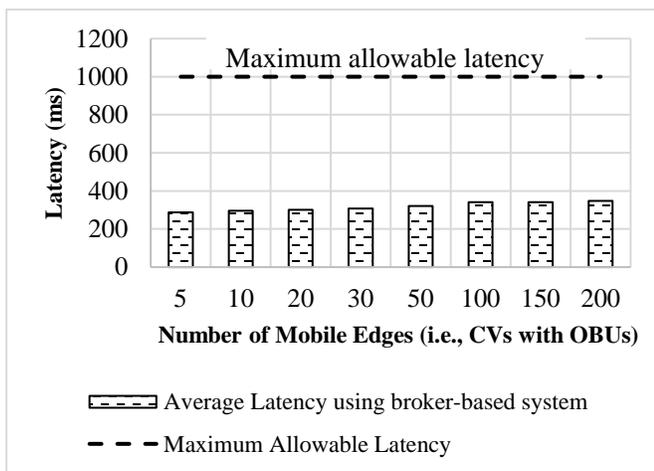

**Figure 15** Comparison of average data delivery latency with different number of mobile edges (i.e., CVs with OBUs) (Scenario 1)

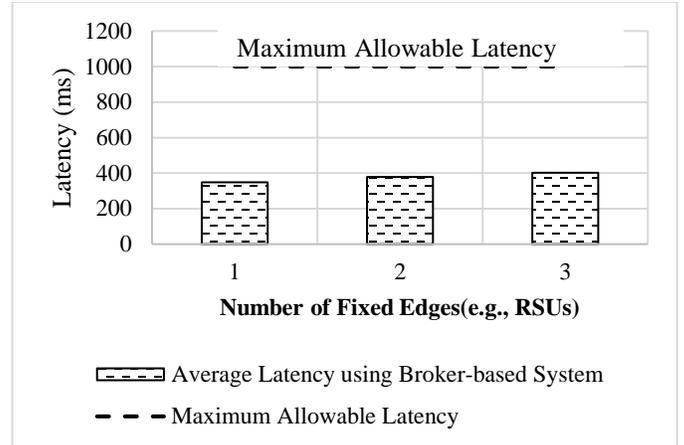

**Figure 16** Comparison of average data delivery latency with different number of fixed edges (i.e., RSUs) (Scenario 2)

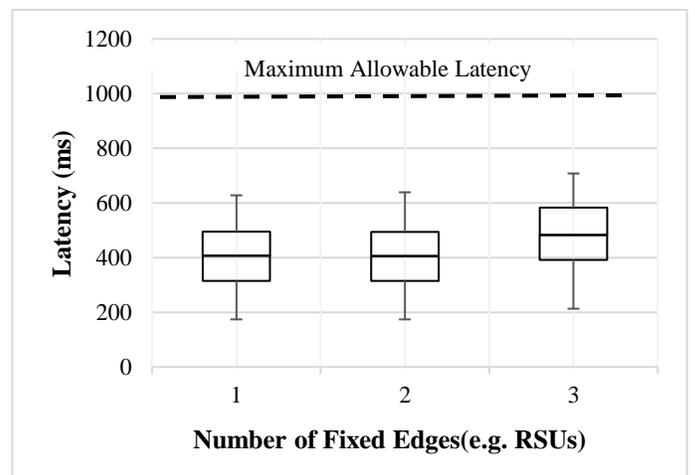

**Figure 17** Latency distribution for varying number of fixed edges (Scenario 2)

Thus, the broker-based system can maintain a lower message delivery latency irrespective of the number of producers and consumers thus ensuring scalability. In our experiment, we have used default configuration of a Kafka broker-based system (e.g., replication factor =1, topic partition =1, and single broker). However, the configuration (e.g., topic partitions, replication, multiple Brokers) of Kafka Broker-based system can be configured easily to reduce the latency if the latency is higher than the CV application threshold. In addition, by adding additional data management brokers, as presented by Du et al. (2017), CVDeP can be scaled up to receive and share data from additional connected data sources (e.g., personal handheld devices, news media and weather stations, traffic operators).

In our previous analysis, we have considered only the average value of latency to justify the CV application requirements. However, we need to evaluate the reliability of data delivery latency by analyzing the data sample space of latency measurements. For each scenario from Table 4, we have collected on average 6000 samples of latency measurement, and we have ran each scenario four times. Figure 17 shows the box and whisker plot, which shows the value of maximum, median, first quartile, and third quartile latency of the latency measurement data set from which the average latencies are measured for Scenario 2. Figure 17 shows that broker-based system delivers the data within required latency constraint irrespective of the number of fixed edges. All of the sampled data for the broker-based system is below the latency requirement.

### 6.3.4 Field Validation

We have validated the developed 'Traffic Data Collection for Traffic Operations' application in real-world. We have used five mobile edges in our evaluation in the field experiment. Table 5 shows the summary of latency when we developed the application in the CVDeP emulated environment and CU-CAVT. We observed higher latency (maximum, average and minimum) in the field than in the CVDeP. In the field experiment, the data exchange through DSRC between the mobile edge and fixed edge in the field was effected by the environmental inferences, such as trees, roadway slope, and curvature. This causes the higher variation in latency in the field than in the CVDeP. However, latency observed in the field was still far below the latency requirements for mobility applications.

**Table 5**

Summary of latency for 'Traffic Data Collection for Traffic Operations' application using CVDeP and in CU-CAVT with five CVs

| Evaluation Parameter | Latency | | Latency requirements for Mobility Application (Fehr et al., 2014) |
| | Evaluation using CVDeP | Evaluation in the Field | |
|---|---|---|---|
| Maximum Latency | 115 ms | 267 ms | |
| Average Latency | 65 ms | 69 ms | ≤1000 ms |
| Minimum Latency | 4 ms | 6 ms | |

## 7   CONCLUSIONS AND FUTURE WORK

CV technology holds the promise of improving safety and efficiency of traffic operations. To materialize the CV benefits, the active participation of CV researchers and developers is necessary. This can be hindered due to the lack of real-world application development platform that uses real-world and real-time CV data to support the CV application development process including testing and debugging. Our research and development contributes directly by developing a CV application development platform, CVDeP, for an edge-centric CPS. Using this CVDeP, CV application developers can interact with a real-world edge devices, and develop, test and debug CV safety and mobility applications using real-time data. CVDeP is evaluated in terms of latency and throughput for CV applications while enabling the scalability and security of both the platform and application. The evaluation has been performed through two connected vehicle applications: (1) "Forward Collision Avoidance" and; (2) "Traffic Data Collection for Traffic Operations". In the security module, security policies are implemented to maintain the application security and access control to the platform. A flow-based application security is used to monitor the data flow for the CV applications to ensure application security. On the other hand, an authentication procedure is implemented to restrict the access to the edge-centric CPS by unauthorized users. However, we have not evaluated the performance of CVDeP under various cyberattacks (e.g. Denial of Service, Message Alternation, and Impersonation) as security policies for detecting cyber-attacks and related countermeasures are not the focus of our current study.

From our case study, it is revealed that the applications developed using CVDeP are able to fulfill the CV safety and mobility application latency requirements and provide high throughput both for an increasing number of mobile edges, and multiple fixed edges. We showed that, "Forward Collision Avoidance" application' (a CV safety application) developed using CVDeP can fulfill the minimum latency requirement of 200 milliseconds. Also, "Traffic Data Collection for Traffic Operations" application (a CV mobility application) developed using CVDeP with broker-based system shows about 400 milliseconds of latency with 3 fixed edges and 600 mobile edges, which is much lower than the minimum latency requirement of mobility applications (approximately 1000 milliseconds). This also proves the scalability of our CVDeP while fulfilling the latency requirement of CV applications for an edge-centric CPS.

In our evaluation of CVDeP, there exist few limitations such as the resiliency and fault tolerance of the platform are not evaluated. This research is conducted using multiple mobile edges (CVs) and fixed edges (RSUs), and the evaluation is conducted with two CV applications only. In addition, only one system edge is used for our evaluation and only data from mobile and fixed edges are collected to evaluate CVDeP, not the data from other sensors or roadside infrastructure (e.g. Traffic singal conttollers). As CVDeP is being developed and refined further, future study shall include: i) incorporation of data from other traditional data sources (e.g., traffic signal, loop detector) and non-traditional data sources (e.g., news media, weather sensors, social networking sites); ii) evaluation of the fault tolerance and resiliency of the platform; iii) evaluation of multiple applications running simultaneously in multiple system edges while merging information from diverse data sources for a large network (i.e., data residing at local or city/county level, regional or state level, and/or national level); and iv) strategy identification to make the system more secure by incorporating different security threat detection and protection mechanisms against different malicious activity including cyber-attacks.

## ACKNOWLEDGEMENTS

This study is based upon work supported by the USDOT Connected Multimodal Mobility University Transportation Center (C$^2$M$^2$) (Tier 1 University Transpiration Center) headquartered at Clemson University, Clemson, South Carolina, USA. Any opinions, findings, and conclusions or recommendations expressed in this material are those of the author(s) and do not necessarily reflect the views of the USDOT Center for Connected Multimodal Mobility (C$^2$M$^2$), and the U.S. Government assumes no liability for the contents or use thereof. Also, the authors would like to thank Aniqa Chowdhury for editing the paper.

## REFERENCES

ActiveMQ (2017), Available at: https://activemq.apache.org/, accessed July 2017.

Ahmed-Zaid, F., Bai, F., Bai, S., Basnayake, C., Bellur, B., Brovold, S., Brown, G., Caminiti, L., Cunningham, D., Elzein, H., Hong, K., Ivan, J., Jiang, D., Kenney, J., Krishnan, H., Lovell, J., Maile, M., Masselink, D., McGlohon, E., Mudalige, P., Popov, (2011), Vehicle Safety Communications–Applications (VSC-A) Final Report: Appendix Volume 3 Security (No. HS-811 492D).

Agmon, N., & Ahituv, N. (1987), Assessing data reliability in an information system, Journal of Management Information Systems, 4(2), 34-44.

Apache Kafka (2017), Available at: https://kafka.apache.org/, Accessed July 2017.

ARC-IT (Architecture Reference for Cooperative and Intelligent Transportation) (2018), Available at: https://local.iteris.com/arc-it/index.html, accessed January 2018.

Baker, E. H., D. Crusius, M. Fischer, W. Gerling, K. Gnanaserakan, H. Kerstan, F. Kuhnert, J. Kusber, J. Mohs, M. Schule, J. Seyfferth, J. Stephan, and T. Warnke, (2016), Connected Car Report 2016: Opportunities, Risk, and Turmoil on the Road to Autonomous Vehicles.

Balke, K., Charara, H., & Sunkari, S. (2014), Report on Dynamic Speed Harmonization and Queue Warning Algorithm Design.

Chowdhury, M., Rahman, M., Rayamajhi, A., Khan, S., Islam, M., Khan, Z., & Martin, J. (2018), Lessons Learned from the Real-World Deployment of a Connected Vehicle Testbed, Transportation. Research. Board 97th Annual. Proceedings.

Dey, K., Rayamajhi, A., Chowdhury, M., Bhavsar, P., & Martin, J. (2016), Vehicle-to-vehicle (V2V) and vehicle-to-infrastructure (V2I) communication in a heterogeneous wireless network – Performance evaluation, Transportation Research Part C Emering. Technologies.

Du, Y., Chowdhury, M., Rahman, M., Dey, K., Apon, A., Luckow, A., & Ngo, L. B. (2017), A Distributed Message Delivery Infrastructure for Connected Vehicle Technology Applications. IEEE Transactions on Intelligent Transportation Systems. DOI: 10.1109/TITS.2017.2701799.

El Zarki, M., Mehrotra, S., Tsudik, G., & Venkatasubramanian, N. (2015), Security Issues in a Future Vehicular Network Symp. A Q. J. Mod. Foreign Lit.

Fehr, W. (2018), Southeast Michigan Test Bed: 2014 Concept of Operations. https://local.iteris.com/cvria/docs/SE_Michigan_Test_Bed_2014_ConOps_-_v1_-_2014-Dec_29.pdf, Accessed Feb 14, 2018

Felt, A.P., Ha, E., Egelman, S., Haney, A., Chin, E., & Wagner, D., (2012), Android permissions Demystified proceedings eighth Symposium. Usable Privacy Security - SOUPS '12.

Fernandes, E., Paupore, J., Rahmati, A., Simionato, D., Conti, M., & Prakash,A. (2016), FlowFence: Practical Data Protection for Emerging IoT Application Frameworks, Usenix Secur.

Grewe, D.,Wagner, M. & Kutscher, D. (2017), Information-Centric Mobile Edge Computing for Connected Vehicle Environments : Challenges and Research Directions

Intelligent Transportation Systems, (2017), Real-Time Data Capture and Management. Available at: https://www.its.dot.gov/factsheets/realtime_dcm_factsheet.htm, Accessed July 2017.

Islam, M., Chowdhury, M., Li, H., and Hu. H, (2018) Cybersecurity Attacks in Vehicle to Infrastructure Applications and their Prevention," Accepted at Transportation. Research. Board 97th Annual. Proceedings.

Karagiannis, G., Altintas, O., Ekici, E., Heijenk, G., Jarupan, B., Lin, K., & Weil, T. (2011), Vehicular networking: A survey and tutorial on requirements, architectures, challenges, standards and solutions. IEEE communications surveys & tutorials, 13(4), 584-616.

Kenney, J. B. (2011), Dedicated short-range communications (DSRC) standards in the United States. Proceedings of the IEEE, 99(7), 1162-1182.

Kreps, J, Narkhede, H., and Rao, J., (2011), Kafka: a Distributed Messaging System for Log Processing," ACM SIGMOD Work. Netw. Meets Databases, p. 6.

Lopez , P.G., Montresor, A.,Epema, D., Iamnitchi,A., Felber, P., and Riviere, E., (2015), Edge-centric Computing : Vision and Challenges, vol. 45, no. 5, pp. 37–42.

Open Data Portal (2018), Available: https://www.its.dot.gov/data/. Accessed: Feb 2018.

Park. Y., and Kim, H., (2012), Application-level frequency control of periodic safety messages in the IEEE WAVE," IEEE Trans. Veh. Technol., vol. 61, no. 4, pp. 1854–1862.

Paupore, J., Ferdandes, E., Prakash, A., Roy, S., & Ou, X. (2015), Practical always-on taint tracking on mobile devices. In USENIX Workshop on Hot Topics in Operating Systems (HotOS).

Qian, Y., & Moayeri, N. (2008), Design of secure and application-oriented VANETs. In Vehicular Technology Conference, 2008. VTC Spring 2008. IEEE (pp. 2794-2799). IEEE.

Raw. R. S., Kumar, M., & Singh. N., (2013), Security Challenges, Issues and Their Solutions for VANET," International Journal of. Network and Security & Its

Applications.

Rayamajhi, A., Rahman, M., Kaur, M., Liu, J., Chowdhury, M. & Hu, H. (2017), ThinGs in a Fog : System Illustration with Connected Vehicles.

Russello, G., Conti, M., Crispo, B., & Fernandes, E. Moses (2012), Supporting operation modes on smartphones. In ACM Symposium on Access Control Models and Technologies (SAC- MAT).

RabbitMQ, (2017),– Messaging That Just Works, Available at: http://www.rabbitmq.com/, accessed July 2017.

Redis. (2017), Available at: http://redis.io/topics/pubsub, accessed July 2017.

Shelby, Z., & Bormann, C. (2011), 6LoWPAN: The wireless embedded Internet (Vol. 43). John Wiley & Sons.

Sarwar, G., Mehani, O., Boreli, R., & Kaafar, M. A (2013), On the effectiveness of dynamic taint analysis for protecting against private information leaks on android-based devices. In International Conference on Security and Cryptography (SE- CRYPT).

SUMO, (2018), Simulaiton of Urban Mobility http://sumo.dlr.de/wiki/Simulation_of_Urban_MObility_-_Wiki, Accessed. Jan. 7, 2018

USDOT (2018), The Operational Data Environment. https://github.com/usdot-jpo-ode/jpo-ode, Accessed. Feb. 13, 2018

USDOT (2013), 5.9 GHz DSRC Connected Vehicles for Intelligent Transportation Systems.

Wang, P., Yu, G., Wu, X., He, X., and Wang, Y., (2018), Spreading Patterns of Malicious Information on Platooned Traffic in a Connected Environment, Computer Aided Civil Infrastructure Engineering, vol. 0, pp. 1–18.

Websocket, (2017), Available at: https://www.Websocket.org/, accessed July 2017.

Weil, T., (2017), VPKI Hits the Highway: Secure Communication for the Connected Vehicle Program, IT Prof., vol. 19, no. 1, pp. 59–63.

Whyte, W., Weimerskirch, A., Kumar, V., and Hehn, T., (2013), A security credential management system for V2V communications, IEEE Vehicle. Networking. Conference VNC, no. December, pp. 1–8.

Xiang X, W. Qin, and B. Xiang, (2014), Research on a DSRC-based rear-end collision warning model, IEEE Trans. Intell. Transp. Syst., vol. 15, no. 3, pp. 1054–1065.

Yasumoto, K., Yamaguchi, H., & Shigeno, H. (2016), Survey of real-time processing technologies of IoT data streams. Journal of Information Processing, 24(2), 195-202.

Zhu, F., and Ukkusuri, S. V. (2018), Modeling the Proactive Driving Behavior of Connected Vehicles: A Cell-Based Simulation Approach, Computer Aided Civil Infrastructure Engineering, vol. 33, no. 4, pp. 262–281.